\documentclass[aps,prd,onecolumn, groupedaddress, showpacs]{revtex4} 

\usepackage{graphicx}

\begin{document}

\title{Formation of primordial black holes from non-Gaussian \\ perturbations produced in a waterfall transition}

\author{Edgar Bugaev}
\email[e-mail: ]{bugaev@pcbai10.inr.ruhep.ru}

\author{Peter Klimai}
\email[e-mail: ]{pklimai@gmail.com}
\affiliation{Institute for Nuclear Research, Russian Academy of
Sciences, 60th October Anniversary Prospect 7a, 117312 Moscow, Russia}




\begin{abstract}

We consider the process of primordial black hole (PBH) formation originated
from primordial curvature perturbations produced during waterfall transition
(with tachyonic instability), at the end of hybrid inflation.
It is known that in such inflation models, rather large values of curvature
perturbation amplitudes can be reached, which can potentially cause a
significant PBH production in the early Universe. The probability distributions
of density perturbation amplitudes in this case can be strongly non-Gaussian,
which requires a special treatment. We calculated PBH abundances and PBH mass
spectra for the model, and analyzed their dependence on model parameters.
We obtained the constraints on the parameters of the inflationary potential,
using the available limits on $\beta_{PBH}$.

\end{abstract}


\maketitle

\section{Introduction}
\label{sec-intro}

According to the observational data (see, e.g., \cite{Komatsu:2010fb}),
the primordial curvature perturbation $\zeta$ is
Gaussian with an almost scale-independent power spectrum. It means
that the structure of the Universe originated from near-scale invariant
and almost Gaussian fluctuations.
As is well known, in models of slow-roll inflation with one scalar field
the curvature perturbation originates from the vacuum fluctuations during
inflationary expansion, and these fluctuations lead just to practically Gaussian
classical perturbations with an almost flat power spectrum near the time of horizon exit,
in a full agreement with the data. So far, there is a weak
indication of primordial non-Gaussianity (at $(2 - 3) \sigma$ level) from the
Cosmic Microwave Background (CMB)
temperature information from the WMAP 3-, 5-, 7-year data \cite{Yadav:2010fz, Yadav:2007yy}.

It has been pointed out long ago that for inflation with multiple scalar fields
possibilities exist for non-Gaussian fluctuations \cite{Salopek:1988qh, Salopek:1991jq, Fan:1992wv}.
In particular, authors of \cite{Fan:1992wv} had elaborated a model of Cold Dark Matter (CDM)
(motivated by double inflation
scenarios) in which it had been assumed that the initial perturbation field
(the gauge-invariant potential) is a combination of a Gaussian field $\phi_1$
and the square of an another Gaussian field $\phi_2$, $\Phi = \phi_1 + \phi_2^2$,
and, besides, that $\phi_2$ is described by a sharply peaked power spectrum.

The detectable non-Gaussianity is predicted in models with additional scalar fields
contributing to $\zeta$. The time evolution of the curvature perturbation on
superhorizon scales (which is allowed in double inflation \cite{Starobinsky:1986fxa}
and, in general, in multiple-field scenarios) implies that, in principle, a rather
large non-Gaussian signal can be generated during inflation.
The primordial non-Gaussianity of $\zeta$ in multiple-field models can
be calculated using the $\delta N$ approach
\cite{Starobinsky:1982ee, Starobinsky:1986fxa, Sasaki:1995aw, Lyth:2004gb} or
the expression for $\zeta$ through the non-adiabatic pressure perturbation
\cite{Wands:2000dp, Lyth:1998xn, GarciaBellido:1995qq}.
It is important to note that non-Gaussian contributions to $\zeta$ predicted by
all these approaches might be negligible on cosmological scales but rather
large on smaller scales allowing, in principle, primordial black hole (PBH) formation.

There are several types of two-field inflation scenarios in which detectable
non-Gaussianity of the curvature perturbation $\zeta$ can be generated:
curvaton models
\cite{Mollerach:1989hu, Linde:1996gt, Lyth:2001nq, Moroi:2001ct, Lyth:2002my, Enqvist:2001zp},
models with a non-inflaton field causing inhomogeneous reheating
\cite{Dvali:2003em, Kofman:2003nx}, curvaton-type models of preheating
(see, e.g., \cite{Kohri:2009ac} and references therein), models of waterfall transition
that ends the hybrid inflation \cite{Barnaby:2006cq, Barnaby:2006km, Lyth:2010ch,
Gong:2010zf, Fonseca:2010nk, Abolhasani:2010kr, Abolhasani:2010kn, Lyth:2010zq,
Abolhasani:2011yp, Lyth:2011kj, Bugaev:2011qt}.

In these two-field models, the primordial curvature perturbation
has two components: $\zeta_g$, which is a contribution of the inflaton
(almost Gaussian) and $\zeta_\sigma$, which is a contribution of the extra field $\sigma$.
This second component is parameterized by the following way \cite{Boubekeur:2005fj}:
\begin{equation}
\label{eq1}
\zeta_\sigma( {\bf x} ) = a \sigma( {\bf x} ) + \sigma^2( {\bf x} ) - \langle \sigma^2 \rangle.
\end{equation}
If the linear term in (\ref{eq1}) is negligible (i.e., if $a\approx 0$),
one has the ``$\chi^2$-model'', in
which curvature fluctuations are described by $\chi^2$-distribution.
This $\chi^2$-model is a particular case of $\chi_m^2$-model \cite{ng1, ng2, ng3}, i.e., a model
in which the fluctuations generated during inflation have $\chi_m^2$-distributions.
Of course, there is a severe observational constraint on the spectrum
amplitude ${\cal P}_{\zeta_\sigma}$ \cite{Lyth:2006gd} predicted by the $\chi^2$-model,
at cosmological scales. That is,
the dominance of the quadratic term in (\ref{eq1}) and, correspondingly, large
non-Gaussianity are possible only on smaller scales (where, in particular, in
case of a blue tilt the amplitude ${\cal P}_{\zeta_\sigma}$ can be of order of one,
leading to PBH formation).  The possibilities of PBH formation in curvaton-type scenarios
are discussed in \cite{Kohri:2007qn}.

It was shown in the previous work of authors \cite{Bugaev:2011qt} that the power spectrum
of the primordial curvature perturbations from the waterfall field in hybrid inflation
with tachyonic preheating has a form of a broad peak, and the peak value, $k_*$,
depends on the parameters of the inflationary potential, in particular, on the
parameter $\beta$, which is the ratio $|m_\chi^2|/H^2$, where $m_\chi^2$ is
the mass-squared of the waterfall field $\chi$. At small values of $\beta$
($\beta \lesssim 10$) the peak is far beyond horizon ($k_*/aH \ll 1$) and the
perturbations are strongly non-Gaussian (because they have $\chi^2$-distributions
due to the fact that curvature perturbation $\zeta$ depends on the waterfall field
amplitude quadratically, just like in Eq. (\ref{eq1})). It appears that the spectrum
amplitude ${\cal P}_\zeta$ is negligible on cosmological scales but is quite
substantial at small scales (if $\beta \sim 1$) and it is interesting to analyze if
it can be constrained by data of PBH searches.

The effects of non-Gaussian primordial curvature and density perturbations on the
formation of PBHs had been considered in
works \cite{Bullock:1996at, Ivanov:1997ia, PinaAvelino:2005rm, Hidalgo:2007vk, Saito:2008em}.
Our consideration
in the present paper has almost no intersections with the content of
these works. We study in this paper, mostly, two questions: {\it i)} forms of
the PBH mass spectra in non-Gaussian case and {\it ii)} constraints on the inflationary
potential parameters (for the concrete inflation model) following from processes
of PBH formation in radiation dominated era. Note that we consider the case of the
{\it strong} non-Gaussianity (in contrast with, e.g., studies of \cite{Saito:2008em}).

The plan of the paper is as follows. In Sec. \ref{sec-2} we review, very briefly,
main aspects of the model used for a describing of the waterfall transition (the
tachyonic preheating) at the end of hybrid inflation. In particular, we present in this
Section the formula for the curvature perturbation $\zeta$ (derived in our previous
work \cite{Bugaev:2011qt}), which is a basis for all following calculations
of PBH formation. In Sec. \ref{sec-3} we consider problems connected with the process
of PBH formation in radiation era of the early Universe: probability distributions
functions of our model, formula for PBH mass spectrum (following from Press-Schechter
formalism), formula for the relative energy density of the Universe contained in
PBHs. At the end of Sec. \ref{sec-3} we present some illustrative results of PBH
mass spectra calculations. In Sec. \ref{sec-4} we present the resulting constraints
on the parameters of our inflation model following from the studies of PBH formation.
Sec. \ref{sec-concl} contains our conclusions.


\section{The waterfall transition model}
\label{sec-2}

We consider the hybrid inflation model which describes an evolution of the slowly rolling
inflaton field  $\phi$ and the waterfall field $\chi$, with the
potential \cite{Linde:1991km, Linde:1993cn}
\begin{equation}
\label{pot}
V(\phi, \chi)= \left( M^2 - \frac{\sqrt{\lambda}}{2} \chi^2 \right)^2 + \frac{1}{2}m^2\phi^2
+ \frac{1}{2} \gamma \phi^2 \chi^2 .
\end{equation}
The first term in Eq. (\ref{pot}) is a potential for the waterfall field $\chi$ with the false
vacuum at $\chi=0$ and true vacuum at $\chi_0^2=2 M^2/\sqrt{\lambda} \equiv v^2$. The
effective mass of the waterfall field in the false vacuum state is given by
\begin{equation}
\label{phic-gamma}
m_\chi^2(\phi) = \gamma \left( \phi^2 - \phi_c^2 \right), \qquad
\phi_c^2 \equiv \frac{2M^2\sqrt{\lambda}}{\gamma} .
\end{equation}
At $\phi^2>\phi_c^2$ the false vacuum is stable, while at $\phi^2<\phi_c^2$ the
effective mass-squared of $\chi$ becomes negative, and there is a tachyonic instability
leading to a rapid growth of $\chi$-modes and eventually to an end of the inflationary
expansion.

The evolution equations for the fields are given by
\begin{eqnarray}
\label{phi-eq}
\ddot \phi + 3 H \dot \phi - \nabla^2 \phi & = & - \phi (m^2 + \gamma \chi^2),
\\
\label{chi-eq-new}
\ddot \chi+ 3 H \dot \chi - \nabla^2 \chi & = &
(2\sqrt{\lambda}M^2 - \gamma \phi^2 - \lambda \chi^2) \chi.
\end{eqnarray}
From Eq. (\ref{chi-eq-new}), one obtains the equation for Fourier modes of $\delta\chi$:
\begin{equation}
\label{deltaddotchik}
\delta \ddot\chi_k + 3 H \delta \dot\chi_k + \left(\frac{k^2}{a^2}-\beta H_c^2 +
\gamma \phi^2 \right) \delta \chi_k = 0 .
\end{equation}
Here, $a=a(t)$ is the scale factor and the parameter $\beta$ is given by the relation
\begin{equation}
\beta = 2 \sqrt{\lambda} \frac{M^2}{H_c^2}.
\end{equation}

The solution of Eq. (\ref{phi-eq}) (in which we ignore
gradient term due to the choice of a uniform $\phi$-gauge) is
(for $t>t_c$, $t_c$ is the critical point when
the tachyonic instability begins)
\begin{equation}
\label{phi-t}
\phi = \phi_c e^{-r H_c (t-t_c)}, \qquad r=\frac{3}{2} - \sqrt {\frac{9}{4}- \frac{m^2}{H_c^2}}.
\end{equation}

The time evolution of $\delta \chi_k$ during the growth era of the waterfall was studied,
numerically, in the previous work of authors \cite{Bugaev:2011qt}. We used in \cite{Bugaev:2011qt}
an artificial cut-off of large-$k$ modes, which corresponds to considering only the waterfall field
modes that already became classical near the beginning of the growth era. The classical
nature of the waterfall field had been discussed in \cite{Lyth:2010zq}, in the approximation
when the expansion of the Universe is negligibly small. It has been shown in \cite{Lyth:2010zq}
that, in the Heisenberg picture of the quantum theory, the operator $\hat{\delta \chi_k}$ has,
at not very large $k$, almost trivial time dependence (during the most part of the growth era).
Namely, $\hat{\delta \chi_k}$ is a constant operator times a $c$-number, which means that the
perturbation is classical (this issue is elaborated in detail in the literature on the
quantum-to-classical transition \cite{Polarski:1995jg, Kiefer:1998qe, Lyth:2005ze}).

Following \cite{Lyth:2010zq} we assume that the waterfall transition ends when the last term
in right-hand side of Eq. (\ref{phi-eq}) becomes equal to the preceding one, i.e, when
\begin{equation}
\label{chinl}
\langle (\delta\chi)^2 \rangle  = \frac{m^2}{\gamma} \equiv \chi^2_{nl} .
\end{equation}

The main equation for a calculation of the primordial curvature
perturbation (on uniform density hypersurfaces) is \cite{Wands:2000dp}
(see also \cite{GarciaBellido:1995qq, Lyth:1998xn})
\begin{equation}
\label{zetapnad}
\zeta = - \int dt \frac{H \delta p_{nad}}{p+\rho},
\end{equation}
where the non-adiabatic pressure perturbation is $\delta p_{nad}=\delta p - c_s^2 \delta\rho$
and the adiabatic sound speed is $c_s^2=\dot p / \dot \rho$. The formula (\ref{zetapnad})
follows from the ``separated universes'' picture
\cite{Starobinsky:1982ee, Starobinsky:1986fxa, Sasaki:1995aw, Wands:2000dp, Lyth:2004gb}
where, after smoothing over sufficiently
large scales, the universe becomes similar to an unperturbed FRW cosmology. In our
case, one has
\begin{equation}
\label{dpnad}
\delta p_{nad} = \delta p_{\chi} - \frac{\dot p}{\dot \rho} \; \delta \rho_{\chi}.
\end{equation}
Energy density $\rho$ and pressure $p$ is a sum of contributions of $\phi$ and $\chi$
fields. Eq. (\ref{dpnad}) takes into account that in $\delta p_{nad}$ there is
no contribution from $\phi$ field.

For the curvature perturbation, we have the formula
\begin{equation}
\label{zeta-minusA}
\zeta = \zeta_\chi = - A (\chi^2 - \langle \chi^2 \rangle)
\end{equation}
where $\chi^2$ and $\langle \chi^2 \rangle$ are determined at the time of an
end of the waterfall, $t = t_{end}$, and $A$ is given by the integral \cite{Bugaev:2011qt}
\begin{equation}
\label{Adefined}
A = \int \limits_0^{t_{end}} \frac{H_c dt}{\dot\phi^2(t) + \langle \dot\chi^2(t) \rangle}
\left( \frac{f(t)} {f(t_{end})} \right)^2 \frac{1}{2}
\left[- m_\chi^2(t) + \left( \frac{\dot f(t)}{f(t)} \right)^2  -
  \frac{\dot p}{\dot \rho}
  \left( m_\chi^2(t) + \left( \frac{\dot f(t)}{f(t)} \right)^2 \right) \right].
\end{equation}
Here, the function $f(t)$ describes the time evolution of
the waterfall field, which is almost independent on $k$ \cite{Lyth:2010zq, Bugaev:2011qt},
\begin{equation}
\chi({\bf x}, t ) = C ({\bf x}) f(t) .
\end{equation}

It was shown in \cite{Bugaev:2011qt} that for $\beta \sim 1$, the curvature perturbation
spectrum will reach values of ${\cal P}_\zeta \sim 1$ in a broad interval of other model parameters
(such as $r$, $\gamma$, $H_c$). The peak values, $k_*$, for small $\beta$,
are far beyond horizon, so, the smoothing over the horizon size will
not decrease the peak values of the smoothed spectrum. Furthermore, the spectrum
near peak remains strongly non-Gaussian after the smoothing. The calculations of \cite{Bugaev:2011qt},
based on the quadratic inflaton potential, show that for $\beta \lesssim 100$ and in the broad
interval of $r$ the peak value $k_*$ can be estimated by the simple relation:
\begin{equation}
\frac{k_*}{aH} \sim e^{-N},
\end{equation}
where $N$ is the number of $e$-folds during the waterfall transition. The similar estimate
is contained in the recent work \cite{Lyth:2011kj}. Note that for $k\ll k_*$, we obtain the
well-known (e.g., \cite{Gong:2010zf, Lyth:2010ch, Lyth:2010zq}) result: ${\cal P}_\zeta (k) \sim k^3 $.


\section{PBH production from non-Gaussian perturbation}
\label{sec-3}

\subsection{PBH formation threshold}

A production of PBHs (about these objects, see, e.g., reviews \cite{Khlopov:2008qy, Carr:2009jm})
during reheating process had been considered in works
\cite{Bassett:2000ha, Green:2000he, Suyama:2004mz, Suyama:2006sr, Finelli:2000gi}.
PBH formation in connection to non-Gaussianity has been studied
in \cite{Bullock:1996at, Ivanov:1997ia, PinaAvelino:2005rm, Saito:2008em, Hidalgo:2007vk}.
In all those papers,
the case of rather weak non-Gaussianity has been considered. In the present work, we
study in detail the case of strong non-Gaussianity (i.e., one when quadratic term in Eq. (\ref{eq1})
dominates). Furthermore, we have an opposite sign for the quadratic term due to sign in
Eq. (\ref{zeta-minusA}), which
leads to very different dependencies of PBH abundances on the power spectrum amplitude
compared to the usually considered (Gaussian or almost Gaussian) cases
(such behavior was qualitatively described in \cite{Lyth:2010zq}).

The classical PBH formation criterion in the radiation-dominated epoch is \cite{Carr:1974nx}
\begin{equation}
\label{limit-delta}
\delta > \delta_c \approx 1/3,
\end{equation}
where $\delta$ is the smoothed density contrast at horizon crossing (at this moment, $k=a H$).
The Fourier component of the comoving density perturbation $\delta$ is related to the Fourier
component of the Bardeen potential $\Psi$ as
\begin{equation}
\delta_k = - \frac{2}{3} \left( \frac{k}{aH} \right)^2 \Psi_k.
\end{equation}
For modes in a super-horizon regime, $\Psi_k \approx -(2/3) \zeta_k$,
so (\ref{limit-delta}) can be translated to a limiting value of the curvature perturbation
\cite{Hidalgo:2007vk}, which is
\begin{equation}
\label{zetac075}
\zeta_{c} = \frac{9}{4} \; \delta_c \approx 0.75.
\end{equation}
If we assume somewhat larger PBH formation threshold, $\delta_c\approx 0.45$ (see,
e.g., \cite{Musco:2008hv}), then
\begin{equation}
\label{zetac1}
\zeta_{c} \approx 1.
\end{equation}
We will not insist on the concrete value of the threshold parameter and, in the following, will
consider both values (\ref{zetac075}) and (\ref{zetac1}) as possible ones.

\subsection{Perturbation probability distributions}

The relation between curvature perturbation $\zeta$ and the waterfall field value
is given by Eq. (\ref{zeta-minusA}), or, using $\sigma_\chi^2 = \langle \chi^2 \rangle$,
\begin{equation}
\label{zetaAchi2sigchi}
\zeta = - A (\chi^2 - \sigma_\chi^2) = \zeta_{max} - A \chi^2, \qquad \zeta_{max} \equiv A \sigma_\chi^2.
\end{equation}
Here, $A$ and $\sigma_\chi^2$ generally depend on the smoothing scale $R$.
The distribution of $\chi$ is assumed to be Gaussian, i.e.,
\begin{equation}
\label{pchiGauss}
p_\chi(\chi) = \frac{1}{\sigma_\chi \sqrt{2\pi}} \; e^{-\frac{\chi^2}{2 \sigma_\chi^2} }.
\end{equation}

The distribution of $\zeta$ can be easily obtained from (\ref{zetaAchi2sigchi}, \ref{pchiGauss}):
\begin{equation}
\label{pzetaDISTR}
p_\zeta(\zeta) = p_\chi \left| \frac{d\chi}{d\zeta} \right| = \frac{1}{\sqrt{2\pi \zeta_{max}
(\zeta_{max} - \zeta) } } \;
e^{\frac{\zeta - \zeta_{max}}{2\zeta_{max}} }, \qquad \zeta < \zeta_{max},
\end{equation}
which is just a $\chi^2$-distribution with one degree of freedom, with an opposite sign of the argument,
shifted to a value of $\zeta_{max}$. As required, $\langle \zeta \rangle = 0$ and
\begin{equation}
\label{zeta2avg}
\langle \zeta^2 \rangle = \int \limits_{-\infty}^{\zeta_{max}} \zeta^2 p_\zeta(\zeta) d \zeta
= 2 \zeta_{max}^2.
\end{equation}
On the other hand,
\begin{equation}
\label{zeta2avgP}
\langle \zeta^2 \rangle = \sigma_\zeta^2 = \int {\cal P}_\zeta(k) W^2(kR) \frac{dk}{k},
\end{equation}
where $W(kR)$ is the Fourier transform of the window function, and we use a Gaussian one,
$W^2(kR) = \exp(-k^2 R^2)$, in this work.

From (\ref{zeta2avg}, \ref{zeta2avgP}) we can write for $\zeta_{max}$ (we now denote the
argument $R$ explicitly):
\begin{equation}
\label{zetamaxexpr}
\zeta_{max}(R) = \left[ \frac{1}{2} \int {\cal P}_\zeta(k) W^2(kR) \frac{dk}{k} \right]^{1/2}.
\end{equation}
Everywhere below, we use the following notation:
$\zeta_{max}(R=0) \equiv \zeta_{max}$. So,
\begin{equation}
\label{zetamax0}
\zeta_{max} = \left[ \frac{1}{2} \int {\cal P}_\zeta(k)\frac{dk}{k} \right]^{1/2}= \frac{1}{\sqrt{2}}
\langle \zeta^2 \rangle^{1/2}.
\end{equation}
It is clear that PBHs can be produced in
the early Universe, if $\zeta_{max} > \zeta_c$.

\begin{figure}
\includegraphics[trim = 0 0 0 0, width=8.2 cm]{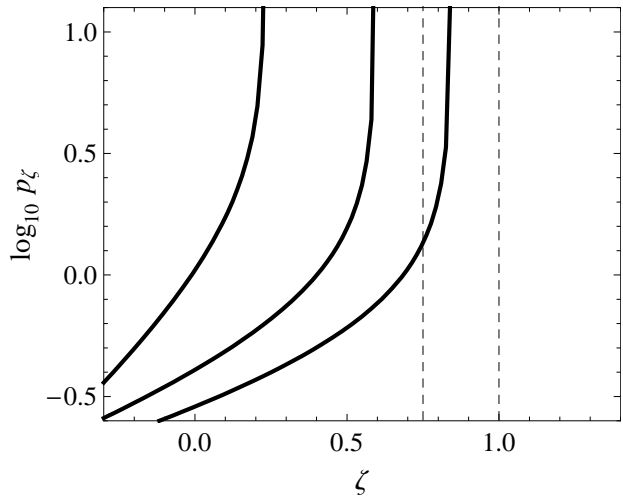} %
\caption{
The probability distribution $p_{\zeta}(\zeta)$ for $\Sigma=0.7$, ${\cal P}_\zeta^0=0.4$
and for different values of $R$: from left to right, $R=10 k_0^{-1}, k_0^{-1}, 0.1 k_0^{-1}$.
The possible (model dependent) values of $\zeta_c$ ($\zeta_c=0.75$ and $\zeta_c=1$) are
shown by dashed lines.
}
\label{fig-pzeta0}
\end{figure}

\subsection{Formulae for PBH mass spectrum and abundance}


In general, a Press-Schechter approach \cite{PS} used for calculations of the number density of clusters
in different scenarios for the formation of structure in the Universe, does not assume a use
of the assumption that the initial perturbations have just Gaussian distributions. For
example, in \cite{Chiu:1997xb} the Gaussianity of the cosmological density field was tested using two
different models for probability distribution functions: a standard Gaussian model and
a texture (see, e.g., \cite{Gooding}) model. In \cite{PinaAvelino:2005rm} the Press-Schechter
formalism had been used for the case when an initial density field has a $\chi^2$-distribution,
and PBH abundances, as a function of black hole mass, for a power-law primordial
power spectrum, were calculated.

The energy density fraction of the Universe contained in collapsed objects of initial mass
larger than $M$
in Press-Schechter formalism \cite{PS} is given by
\begin{equation}
\label{PSformalism}
\frac{1}{\rho_i} \int\limits_M^{\infty} \tilde M n(\tilde M) d \tilde M = \int\limits_{\zeta_{c}}^{\infty}
p_{\zeta}(\zeta) d\zeta = P(\zeta>\zeta_{c}; R(M), t_i),
\end{equation}
where function $P$ in right-hand side is the probability that in the region of comoving
size $R$ the smoothed value of $\zeta$ will be larger than the PBH
formation threshold value, $n(M)$ is the mass spectrum of the collapsed objects,
and $\rho_i$ is the initial energy density. Here we ignore
the dependence of the curvature perturbation $\zeta$ on time after the end of the waterfall,
assuming it does not change in super-horizon regime, until the perturbations enter horizon
at $k=aH$.

The mass spectrum of the collapsed objects, $n(M)$, is given by
\begin{equation}
\label{nMPS}
n(M) = 2 \frac{\rho_i}{M} \left| \frac{\partial P}{\partial R} \right| \frac{dR}{dM},
\end{equation}
where, as usual, the factor 2 approximately takes into account the fact that underdense regions also
collapse. In the above formula, $M$ is the initial fluctuation mass corresponding
to the fluctuation of the comoving scale $R$,
\begin{equation}
\label{MR}
M = \frac{4 \pi}{3} \rho_i (a_i R)^3; \qquad \frac{dR}{dM} =
 (4\pi)^{-1/3} 3^{-2/3} \rho_i^{-1/3} a_i^{-1} M^{-2/3}
\end{equation}
($a_i$ is the value of the scale factor at $t=t_i$, $M$ is calculated at the
moment $t_i$ corresponding to the time of the end of the waterfall, which is assumed
to be close to reheating time; note that comoving fluctuation mass is not constant).

The horizon mass corresponding to the time when fluctuation with initial mass $M$
crosses horizon is (see \cite{Bugaev:2008gw})
\begin{equation}
M_h = M_i^{1/3} M^{2/3},
\end{equation}
where $M_i$ is the horizon mass at the moment $t_i$,
\begin{equation}
\label{Mi}
M_i \approx \frac{4\pi}{3} t_i^3 \rho_i \approx \frac{4\pi}{3} (H_c^{-1})^3 \rho = \frac{4 \pi M_P^2}{H_c}
\end{equation}
(here, we used Friedman equation, $\rho_i = 3 M_P^2 H_c^2$).
The reheating temperature of the Universe is \cite{Bugaev:2008gw}
\begin{equation}
\label{TRH}
T_{RH} = \left( \frac{90 M_P^2 H_c^2} {\pi^2 g_*} \right)^{1/4}, \qquad g_* \approx 100.
\end{equation}

For simplicity, we will use the approximation that mass of the produced black hole is
 proportional to horizon mass, namely,
\begin{equation}
\label{MBH-Mh}
M_{BH} = f_h M_h = f_h M_i^{1/3} M^{2/3},
\end{equation}
where $f_h \approx (1/3)^{1/2} = {\rm const}$ (this particular value of
$f_h$ corresponds to a threshold of the PBH production in Carr-Hawking collapse, see, e.g.,
Appendix of paper \cite{Bugaev:2008gw}).
Our final qualitative conclusions do not depend on the value of $f_h$. In more accurate
analysis, one must take into account that the connection between $M_{BH}$ and $M_h$ is
more complicated, and, in particular, depends on the type of the gravitational collapse
\cite{Carr:1974nx, Niemeyer:1997mt, Musco:2008hv, Bugaev:2008gw}.
Moreover, there can be a dependence on a shape of the radial fluctuation
profile \cite{Hidalgo:2008mv}.

Using Eqs. (\ref{nMPS}, \ref{MR}, \ref{MBH-Mh}), the PBH number density (mass spectrum)
can be written as
\begin{equation}
\label{nBH}
n_{BH}(M_{BH}) = n(M) \frac {dM}{dM_{BH}} = \left( \frac{4 \pi}{3} \right)^{-1/3}
\left| \frac{\partial P}{\partial R }\right|
 \frac{f_h \rho_i^{2/3} M_i^{1/3} } {a_i M_{BH}^2}.
\end{equation}

We can estimate the relative energy density of the Universe contained in PBHs, at the
moment of time $t$ (assuming radiation-dominated Universe with $t<t_{eq}$ and
ignoring the PBH mass change due to accretion or evaporation) as follows
\begin{equation}
\Omega_{PBH}(t) \approx \frac{1}{\rho(t)} \left( \frac{a_i}{a(t)} \right)^3 \int n_{BH} M_{BH} d M_{BH}.
\end{equation}
Using the scaling relations $\rho\sim t^{-2}$, $a\sim t^{1/2}$ and considering the moment
of time for which horizon mass is equal to $M_h$, we obtain
\begin{equation}
\Omega_{PBH}(M_h) \approx \frac{1}{\rho_i} \left( \frac{M_h}{M_i} \right)^{1/2} \int n_{BH} M_{BH}^2 d \ln M_{BH}.
\end{equation}
It is well known that for an almost monochromatic PBH mass spectrum, $\Omega_{PBH}(M_h)$ coincides with
the traditionally used parameter $\beta_{PBH}$ (energy density fraction of the
Universe contained in PBHs at the moment of their formation). Although all PBHs do not form
at the same moment of time, it is convenient to use the combination
$M_i^{-1/2} \rho_i^{-1} M_{BH}^{5/2} n_{BH}(M_{BH})$ to have a feeling of how many PBHs
actually form. We will use this combination in the following (Fig. \ref{fig-nBH} below).

One should note that, strictly speaking, observational constraints on the parameter
$\beta_{PBH}$ are obtained using the assumption that the PBH formation takes place
at a single epoch and the corresponding spectrum ${\cal P}_\zeta$ has a narrow peak
at some value of $k$. In our case, a width of the peak is not very small (see Fig. 4
in \cite{Bugaev:2011qt}). But, as is well known, the black hole abundance is extremely
sensitive to an amplitude of ${\cal P}_\zeta(k)$ and, correspondingly, the bound
on ${\cal P}_\zeta(k)$ weakly depends on the value of $\beta_{PBH}$. Our main
aim is to determine constraints on the inflation model parameters (in particular, on the
value of $\beta$), and, as we show in the present paper, the black hole abundances
depend on $\beta$ and on amplitude of ${\cal P}_\zeta(k)$ very strongly. In such a
situation the form of the ${\cal P}_\zeta$-spectrum in real scenario of PBH
formation (a size of the spectrum's width) is not very essential.

\begin{figure}
\includegraphics[trim = 0 0 0 0, width=8.3 cm]{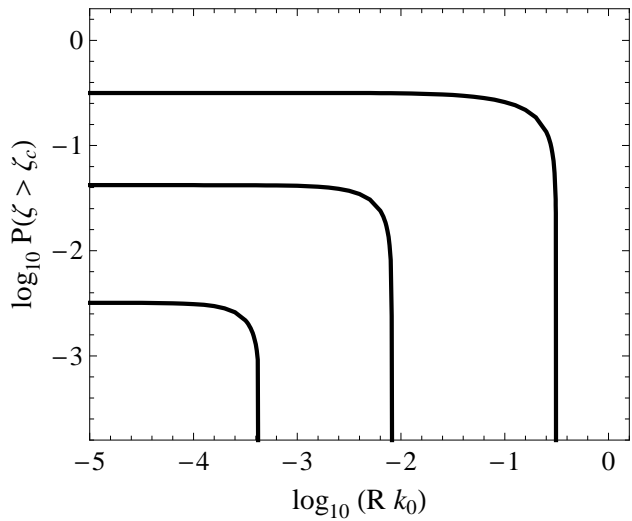} %
\caption{
The probability $P(\zeta > \zeta_c)$ as a function of $R$.
From top to bottom, ${\cal P}_\zeta^0 = 0.4 (\zeta_{max}\approx 0.9),
0.28 (\zeta_{max}\approx 0.752), 0.27846 (\zeta_{max}\approx 0.750012)$.
For all curves, $\Sigma=0.7$, $\zeta_c=0.75$.
}
\label{fig-Plarger}
\end{figure}

\subsection{PBH mass spectrum calculation}

The power spectra ${\cal P}_\zeta(k)$ from waterfall transition process,
for different sets of parameters, have been calculated in
our previous paper \cite{Bugaev:2011qt}. For the purpose of this section, it is convenient to parameterize
the curvature perturbation power spectrum as follows,
\begin{equation}
\label{calP-param}
{\cal P}_\zeta(k) = {\cal P}_\zeta^0 \exp{\left[- \frac{(\lg {k/k_0})^2 } { 2 \Sigma^2 } \right]},
\end{equation}
where ${\cal P}_\zeta^0$ gives the maximum value approached by the spectrum, $k_0$ is
the comoving wave number corresponding to the position of the maximum and $\Sigma$ determines
the width of the spectrum. Evidently, values of the parameter $k_0$ should be equal
to the corresponding peak values, $k_*$, of the ${\cal P}_\zeta(k)$-curves
calculated in \cite{Bugaev:2011qt}.

What are the typical values of the parameters ${\cal P}_\zeta^0$ and $\Sigma$?
It was shown in \cite{Bugaev:2011qt} that
the spectrum ${\cal P}_\zeta(k)$ can reach values of order of $1$ if $\beta\sim 1$.
It follows from Fig. 4 of \cite{Bugaev:2011qt} that for $\beta=2, r=0.1$ one has
$\Sigma\approx 0.7, {\cal P}_\zeta^0 \approx 0.4$ (and $\zeta_{max}\approx 0.9$), while for $\beta=1, r=0.1$
the peak is more wide and high: $\Sigma\approx 1.0, {\cal P}_\zeta^0 \approx 1.2$
(with $\zeta_{max}\approx 1.86$).

So, in case of $\beta=2$ the PBHs are produced if formation threshold is assumed as in (\ref{zetac075}) while
they are not yet produced in case of (\ref{zetac1}).

We note that, generally, the values of ${\cal P}_\zeta^0$, $\Sigma$ and $k_0$ depend on the rather complex
interplay between the potential parameters $\beta$, $r$, $H_I$ and $\gamma$. In particular,
the position of the peak, $k_0$, depends on $H_I$ and the number of $e$-folds which waterfall takes,
and thus varies with the change of any of the above 4 parameters. So, in this Section, we will
consider ${\cal P}_\zeta^0$, $k_0$ and $\Sigma$ as independent parameters.

In Fig. \ref{fig-pzeta0} we show the distribution $p_{\zeta}(\zeta)$
(Eq. (\ref{pzetaDISTR})) for a fixed set of
parameters $\Sigma$, ${\cal P}_\zeta^0$ (physically, they correspond to $\beta=2, r=0.1$ case)
and for different values of the fluctuation size (smoothing scale) $R$. It is seen that in this case,
for larger values of $R$ the fluctuation spectrum is not ``strong'' enough to produce PBHs, while
for smallest $R$ the PBHs will form (if we assume threshold (\ref{zetac075}) - both values
of $\zeta_c$ that we consider are also shown in the Figure).

The probability $P(\zeta > \zeta_c)$ (used in Eq. (\ref{PSformalism})) for fixed values of
$\Sigma$ and $\zeta_c$, but for different ${\cal P}_\zeta^0$ is shown in Fig. \ref{fig-Plarger}.
The value of ${\cal P}_\zeta^0$ was fine tuned for the bottom curve so that
$\zeta_{max}$ does not go below $\zeta_c$ (and $\zeta_{max} - \zeta_c \ll 1$).
Once $\zeta_{max}$ drops below $\zeta_c$, no PBHs ever form (at least for the classical PBH
production scenario which we consider).
For the bottom curve, it happens at $k_0 R \ll 1$ just because $\zeta_{max}(R)-\zeta_c > 0$ only
for the smallest values of $R$ (see Eq. (\ref{zetamaxexpr})).
For each curve in Fig. \ref{fig-Plarger} we see a
sharp drop of $P(\zeta > \zeta_c)$ to the zero at the value of $R$ when $\zeta_{max}(R)$
reaches $\zeta_c$. Technically, the derivative $\partial P / \partial R$ at this point diverges,
which will lead to a characteristic spike in the PBH mass spectrum, according to Eq. (\ref{nBH}).

\begin{figure}
\includegraphics[trim = 0 0 0 0, width=7.8 cm]{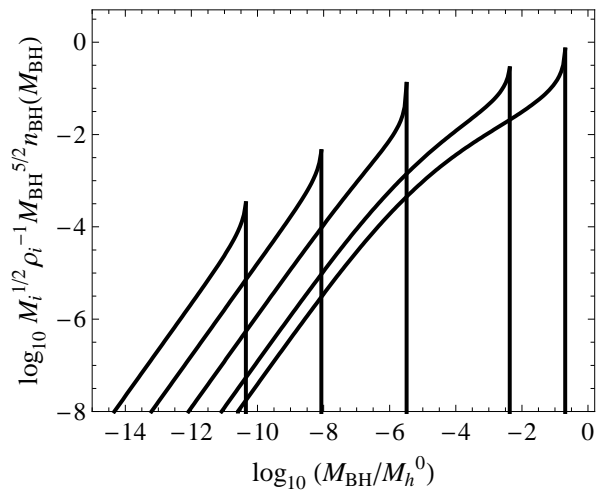} %
\caption{
The PBH mass spectra for different values of the perturbation spectrum
amplitudes. From right to left,
${\cal P}_\zeta^0=1 (\zeta_{max}\approx 1.42), 0.4 (\zeta_{max}\approx 0.9),
0.28 (\zeta_{max}\approx 0.752), 0.27846 (\zeta_{max}\approx 0.750012),
0.2784511 (\zeta_{max}\approx 0.750000068)$.
The position of the peak in ${\cal P}_\zeta(k)$-spectrum is the same for all cases.
For the calculation we used the value $\Sigma=0.7$, and $\zeta_c=0.75$.
The mass $M_h^0$ corresponds to horizon mass at the moment of time when perturbation with
comoving wave number $k_0$ enters horizon.
}
\label{fig-nBH}
\end{figure}

Results of the PBH mass spectra calculations using formula (\ref{nBH}) are shown in
Fig. \ref{fig-nBH}. Again, we are doing some fine-tuning for part of the curves, choosing
parameters so that $\zeta_{max} - \zeta_c \ll 1$. In these cases, such fine tuning allows
us to reach $\beta_{PBH} \sim M_i^{-1/2} \rho_i^{-1} M_{BH}^{5/2} n_{BH}(M_{BH}) \sim 10^{-3}$ or so.
The mass of the produced PBHs, as can be seen from the same Figure, is several
orders below $M_h^0$.

It is well known \cite{Carr:2009jm} that for rather large PBH masses,
say, $M_{BH}\gtrsim 10^{10}\;$g, the constraints
on their abundance from different types of sources (such as gravitational
constraints and constraints following from non-observation of the products of PBH's Hawking
evaporation) are rather severe, $\beta_{PBH} \lesssim (10^{-27} \div 10^{-10})$, or so.
It turns out that in the present model it is possible to reach such small values
of $\beta_{PBH}$ only with extreme fine tuning of inflationary potential
model parameters. However, in the range of masses below $M_{BH} \sim 10^{10}\;$g the constraints
on $\beta_{PBH}$ are not so severe, and in this mass range,
$\beta_{PBH} \sim 10^{-3}$ is not forbidden by the observations. Such light PBHs
evaporate very fastly (long before nucleosynthesis) and the hope of their possible detection
is mainly due to high-frequency GWB which they generate through Hawking evaporation
(see, e.g., \cite{BisnovatyiKogan:2004bk, Anantua:2008am, Dolgov:2011cq}; in these
papers, PBHs with masses $M_{BH}\sim 10^5\;$g and values of $\beta_{PBH} \sim 10^{-3}$ are
considered).

\section{Constraints on the waterfall transition model}
\label{sec-4}

We have seen that in the waterfall model considered, PBH abundance severely depends
on the amplitude of the curvature perturbation spectrum: once $\zeta_{max}$ is above
$\zeta_c$, PBHs are produced intensively. Demanding that PBHs do not form in the early
Universe, we can impose the bound on parameters of the inflaton potential.
From the condition $\zeta_{max} < \zeta_c$ one has, for two
fixed values of $\zeta_c$, the following constraints
($r=0.1$; there is a weak dependence on this parameter
but the result is almost independent on $\gamma, H_c$):
\begin{equation}
\zeta_c=0.75: \;\; \beta>2.3 , \;\; {\cal P}_\zeta^0 < 0.29 ;
\end{equation}
\begin{equation}
\zeta_c=1: \;\; \beta>1.65 , \;\; {\cal P}_\zeta^0 < 0.55 .
\end{equation}
These constraints are not based on the comparison with data on $\beta_{PBH}$.

It is interesting to estimate the mass of PBHs that can be produced by this model,
and corresponding horizon masses.
For the horizon mass corresponding to the peak position \cite{Bugaev:2011qt}
one has the relation
\begin{equation}
\label{Mh0N}
M_h^0 \approx e^{2 N} M_i,
\end{equation}
where $N$ is the number of $e$-folds that waterfall transition takes.
This approximate relation follows from the estimate $k_*/aH\sim e^{-N}$ obtained
in \cite{Bugaev:2011qt}. Here, $k_*$ is the peak value of the $\zeta$ power spectrum,
$a H$ corresponds to an end of the waterfall.
We find that $N$ is
highly parameter-dependent: although ${\cal P}_\zeta^0$ mostly depends on $\beta$ and
rather weakly on other parameters, $N$ depends on $\gamma$ (or $\phi_c$)
in a strong way. Here we estimate the range of $N$ values to get an idea what PBH mass
range is given by Eq. (\ref{Mh0N}).

For the purpose of this estimate, we take the maximal possible value of $\phi_c$ to be
equal to the (reduced) Planck mass $M_P$. The lower limit is obtained from the classicality
condition \cite{Lyth:2010zq}: to have an effective waterfall
(classical regime is reached
for the $\chi$ field), we need $\sqrt{\gamma} \ll 1/\sqrt{\beta}$, or, at least,
$\gamma \sim 1/\beta$ (numerical solutions show that for sets of parameters we consider,
the waterfall is still effective in this case). So we consider the lower limit for
$\phi_c$ to be $\phi_c = \beta H_c$.

We find that $N \approx 4$ for the limit of $\phi_c = \beta H_c$, both for $\beta=1.65$
and $\beta=2.3$; this result turns out to be independent on $H_c$.
With the growth of $\phi_c$, $N$ also increases. For $\phi_c = M_P$ and $H_c=10^{11}\;$GeV,
$N\approx 32$ for $\beta=2.3$ and $N\approx 41$ for $\beta=1.65$ (these values of $N$ also have some
dependence on inflation energy scale $H_c$).

\begin{figure}
\includegraphics[trim = 0 0 0 0, width=8.6 cm]{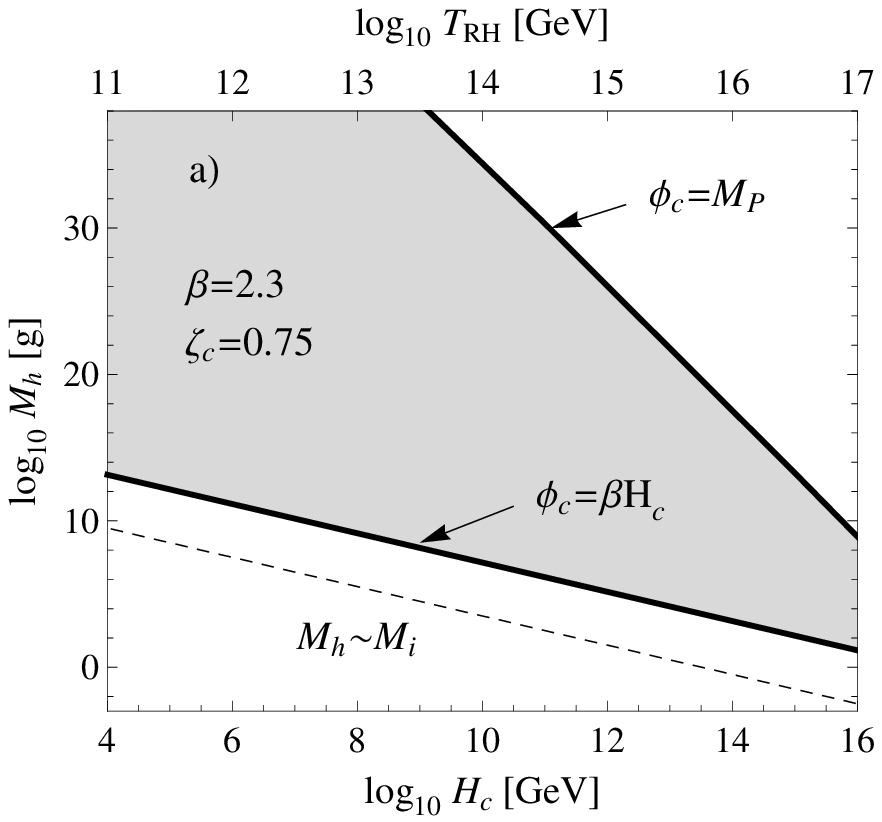} $\;\;\;$ %
\includegraphics[trim = 0 0 0 0, width=8.6 cm]{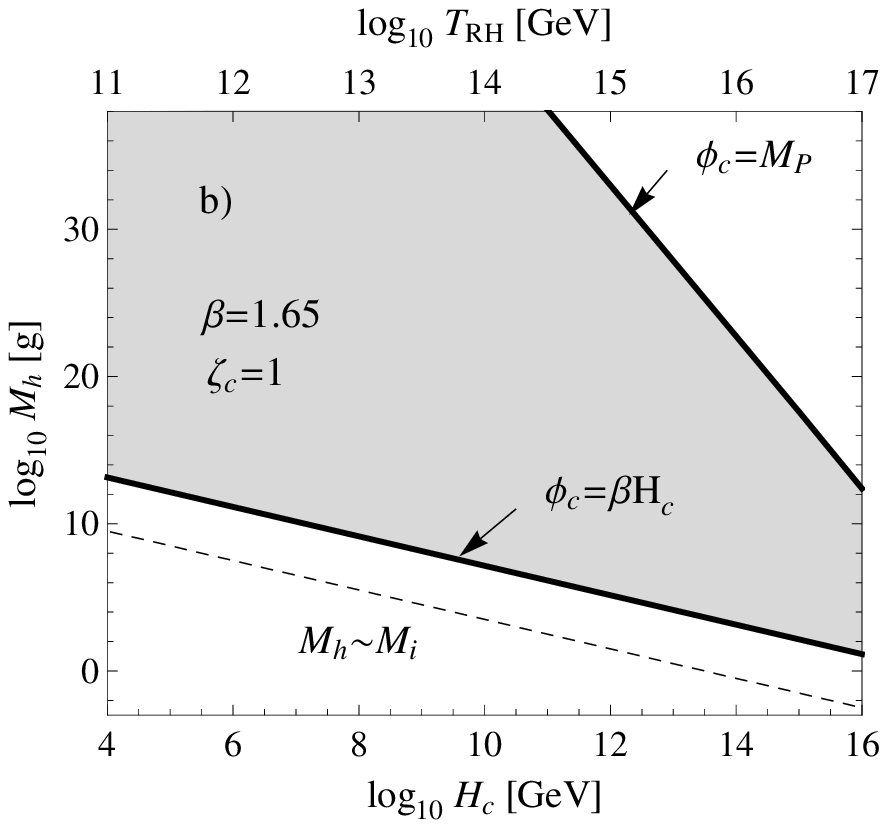} %
\caption{
The horizon mass regions corresponding to the position of the peak in curvature
perturbation power spectrum (shaded areas).
{\bf a)}$\; \beta\approx 2.3, \zeta_c=0.75$; {\bf b)} $\beta\approx 1.65, \zeta_c=1$.
In both cases, the value of $\beta$ is just enough to produce PBHs, and we vary
the parameter $\phi_c$ (or, equivalently, $\gamma$) between minimal and maximal possible values.
The exact PBH abundance and relation of characteristic PBH mass $M_{BH}$ to $M_h$ will depend
on values of parameters in a fine-tuning regime (see Fig. \ref{fig-nBH} for an illustration).
}
\label{fig-masses}
\end{figure}

We plot the possible regions of $M_h$ in Fig. \ref{fig-masses}. It is seen that the possible
mass range is very wide. As discussed above, the exact PBH abundance ($\beta_{PBH}$) and relation of
characteristic PBH mass $M_{BH}$ to $M_h$ will depend
on values of parameters in a fine-tuning regime (see Fig. \ref{fig-nBH} for an illustration).

\section{Conclusions}
\label{sec-concl}

We have considered PBH production from strongly non-Gaussian density (curvature) perturbations
in the radiation-dominated era of the early Universe. The main physical model that we
used is the model of hybrid inflation waterfall (with tachyonic preheating),
however, the results may be applied to
different models that produce similar perturbations.

We have given the expressions for perturbation probability distributions and,
on the basis of Press-Schechter formalism, calculated PBH abundances and PBH mass spectra
for the model. The most important result of the paper is that we obtained
limits on the parameters of the potential of our hybrid inflation model.
In particular, we have shown that the parameter $\beta$, which is the ratio
$|m_\chi^2|/H^2$, is limited from below, i.e., $\beta$ is larger than
some value (otherwise the abundance of PBHs will be too high, in contradiction
with available constraints on $\beta_{PBH}$). We
have shown also that these limits on $\beta$ are sensitive to the PBH formation threshold
parameter (in our case, $\zeta_c$). Note that the characteristic PBH masses
that can be (in principle) produced by this model, are shown to depend significantly
on the coupling parameter, $\gamma$, of the inflaton potential. The second important
result is that for our inflation model the possible horizon and
PBH mass regions, corresponding to the peak in perturbation power spectrum, are constrained
(Fig. \ref{fig-masses}).

It was shown also, that to obtain values of PBH abundance not contradicting
with the available limits on $\beta_{PBH}$, for $M_{BH}\gtrsim 10^{10}\;$g, extreme
fine-tuning of the model parameters are needed. Note that the model allows to
more or less naturally (with a degree of fine-tuning similar to one needed
in some single-field inflation models \cite{Saito:2008em, Bugaev:2008bi})
produce PBHs with $M_{BH}\sim 10^5\;$g or so,
and $\beta_{PBH}\sim 10^{-3}$ or so. Such PBHs are one of the possible sources
of high-frequency GW background, which can be observed in future.

Throughout the work, we used simplified gravitational collapse model (using Eq. (\ref{MBH-Mh})),
close to the standard one, to treat PBH formation. The inclusion of critical
collapse \cite{Niemeyer:1997mt, Musco:2008hv}
effects in our calculation would require the replacement of $f_h$ in Eq. (\ref{MBH-Mh}) with a function
proportional to $(\zeta-\zeta_c)^{\gamma_c}$, $\gamma_c$ is around $0.3 - 0.4$. This will
change the results for PBH mass spectrum (Fig. \ref{fig-nBH}) - in particular, the curves
on that Figure will be shifted by $1-2$ orders of magnitude to the left and
their shape will change, but this will not affect other results of the paper.

\bigskip

{\it Note added.} After this work has been published as an e-print, the paper \cite{Lyth:2012yp}
appeared, in which an analytical framework for calculating the curvature perturbation
spectra produced by the hybrid inflation waterfall is developed, in a general case of
any inflaton potential form including the case of $N \gtrsim 1$.

The author of \cite{Lyth:2012yp} finds good agreement between numerical calculations of
our work \cite{Bugaev:2011qt} and his analytic estimates. He also discusses in detail
all assumptions that are made in such models, in particular, he stresses that
the gradient of $\chi$ must be negligible for the calculation using formula (\ref{zetapnad})
to be viable. We proved the smallness of this gradient in \cite{Bugaev:2011qt} for our case (see
Eq. (3.9) of that work). Also, \cite{Lyth:2012yp} discusses the compatibility
of the evolution equation for $\chi$ (our Eq. (\ref{deltaddotchik})) with the
energy continuity equation. The possible inconsistency between these two equations
is due to the fact that interaction term, $\sim \phi \chi^2$, is dropped in
Eq. (\ref{phi-eq}) but term $\sim \phi^2\chi$ is still used in Eqs. (\ref{chi-eq-new}, \ref{deltaddotchik}).
The condition for using this approximation correctly is derived to be \cite{Lyth:2012yp}
\begin{equation}
\epsilon \equiv \frac{1}{|m_\chi(t)| H} \frac{d |m_\chi(t)|}{dt} \ll 1.
\end{equation}
In our case, from Eqs. (\ref{phic-gamma}) and (\ref{phi-t}), one has
\begin{equation}
\label{epsilon}
\epsilon  = \frac{r}{e^{2 r H t} - 1},
\end{equation}
which is small only for $N = Ht \gtrsim 1$. For smaller $H t$, when the waterfall just started,
$\chi_k$ is small, and term proportional to $\phi \chi^2$ in Eq. (\ref{phi-eq}) can be
dropped just due to smallness of $\chi$.
As one can see from Eq. (\ref{epsilon}), the consistency condition depends only on a value
of the parameter $r$ which we fixed throughout the present work. Namely, we used the
value $r=0.1$ and, in this case, the consistency condition is satisfied.


\end{document}